\documentclass[english,12pt]{article}
\usepackage[cp1251]{inputenc}
\usepackage{babel}
\usepackage{amsfonts, amsmath}

\newcommand{\be}{\begin{equation}} \newcommand{\ee}{\end{equation}}

\begin{document}
\title{Quantum Mechanics of the Early Universe and its
Limiting  Transition} \thispagestyle{empty}

\author{A.E.Shalyt-Margolin\hspace{1.5mm}\thanks
{Phone (+375) 172 883438; e-mail: a.shalyt@mail.ru; alexm@hep.by
}, J.G.Suarez
\thanks{Phone (+375) 172 883438; e-mail: suarez@hep.by, jsuarez@mail.tut.by}
}
\date{}
\maketitle
 \vspace{-25pt}
{\footnotesize\noindent  National Center of Particles and High
Energy Physics, Bogdanovich Str. 153, Minsk 220040, Belarus\\
{\ttfamily{\footnotesize
}}

\rm\normalsize \vspace{0.5cm}
\begin{abstract}
In this paper Quantum Mechanics with Fundamental Length is chosen
as the theory for describing the early Universe. This is possible
due to the presence in the theory of General Uncertainty Relations
from which unavoidable it follows that in nature
 a fundamental length exits. Here Quantum
 Mechanics with Fundamental Length is
obtained as a deformation of Quantum Mechanics. The distinguishing
feature of the proposed in this paper approach in comparison with
previous ones, lies on the fact that here density matrix subjects
to deformation as well as so far commutators had been deformed.
The deformed density matrix mentioned above, is named throughout
this paper density pro-matrix. Within our approach two main
features of Quantum Mechanics are conserved: the probabilistic
interpretation of the theory and exact predefined measuring
procedure corresponding to that interpretation. The proposed here
approach allows to describe dynamics. In particular, the explicit
form of deformed Liouville's equation and  the deformed
Shr\"odinger's picture are given. Some implications of obtained
results are discussed. In particular, the problem of singularity,
the hypothesis of cosmic censorship, a possible improvement of the
statistical entropy definition and the problem of information loss
in black holes are considered.
\end{abstract}

\section{Introduction}
In this paper Quantum Mechanics with Fundamental Length (QMFL) is
considered as Quantum Mechanics (QM) of the early Universe. The
main motivation for this choice  is the presence in the theory of
 General Uncertainty Relations (GUR) appropriated to describe the
  behavior of the early Universe
and unavoidable conducting to the concept of fundamental length.
Here QMFL is obtained as a deformation of QM, choosing the
quantity $\beta=l_{min}^{2}/x^{2}$ (where $x$ is the scale) as the
parameter of deformation of the theory. The main difference
between presented
 approach and previous ones lies on the fact, that we propose a
density matrix deformation, as well as so far commutator's
deformation had been proposed. Obtained in such a way density
matrix (generalized density matrix) is called here and throughout
this paper density pro-matrix. Within our approach two very
important features of QM have been conserved. Namely,
 the probability interpretation and exact
predefined measurement procedure, corresponding to this
interpretation have been transferred to QMFL. It was shown that in
the paradigm of expanded universe model there are two different
(unitary non-equivalent) Quantum Mechanics: the first one named
QMFL is describing nature on Planck's scale or on the early
Universe and it is based on GUR. The second one named
 QM and representing passage to the limit from Planck's to low
energy scale is based on Heisenberg's Uncertainty Relations (UR).
Consequently, some well-known quantum mechanical concepts could
appear only in the low energy limit. Further, within the proposed
approach some dynamical aspects of QMFL are described. In
particular, a deformation of the Liouville's equation, the
Shr\"odinger's picture in QMFL as well as some implications of
obtained results are presented. Mentioned implications deal with
the problem of singularity,
 the hypothesis of cosmic censorship, a possible improvement of
the statistical
 entropy definition and also with the problem of information
loss in black holes.

\section{Fundamental Length and Density Matrix}

Using different approaches (String Theory \cite{r2}, Gravitation
\cite{r3}, Quantum Theory of black holes \cite{r4} , etc.) the
authors of numerous papers issued over the last 14-15 years have
pointed out that Heisenberg's Uncertainty Relations should be
modified. Specifically, a high energy correction has to appear
\begin{equation}\label{U2}
\triangle x\geq\frac{\hbar}{\triangle p}+\alpha
L_{p}^2\frac{\triangle p}{\hbar}.
\end{equation}
\noindent Here $L_{p}$ is the Planck's length:
$L_{p}=\sqrt\frac{G\hbar}{c^3}\simeq1,6\;10^{-35}\;m$ and
 $\alpha > 0$ is a constant. In \cite{r3} it was shown
that this constant may be chosen equal to 1. However, here we will use
$\alpha$ as an arbitrary constant without giving it any definite
value.
The  inequality (\ref{U2}) is quadratic in
$\triangle p$:
\begin{equation}\label{U3}
\alpha L_{p}^2({\triangle p})^2-\hbar \triangle x \triangle p+
\hbar^2 \leq0,
\end{equation}
from whence the fundamental length is
\begin{equation}\label{U4}
\triangle x_{min}=2\sqrt\alpha L_{p}.
\end{equation}
Since in what follows we proceed only from the existence of
fundamental length, it should be noted that this fact was
established apart from GUR as well. For instance, from an ideal
experiment associated with Gravitational Field and Quantum
Mechanics a lower bound on minimal length was obtained \cite{r6},
\cite{r7} and  improved in \cite{r8} without using GUR to an
estimate of the form $\sim L_{p}$. \noindent Consider equation
(\ref{U4}) in some detail. Squaring both sides of the equation, we
obtain
\begin{equation}\label{U5}
(\overline{\Delta\widehat{X}^{2}})\geq 4\alpha L_{p}^{2 },
\end{equation}
Or in terms of density matrix
\begin{equation}\label{U6}
Sp[(\rho \widehat{X}^2)-Sp^2(\rho \widehat{X}) ]\geq 4\alpha
L_{p}^{2 }=l^{2}_{min}>0,
\end{equation}
where $\widehat{X}$ is the coordinate operator. Expression
(\ref{U6}) gives the measurement rule used in QM. However, in the
case considered here, in comparison with QM, the right part of
(\ref{U6}) cannot be done arbitrarily near to zero since it is
limited by $l^{2}_{min}>0$ where due to GUR $l_{min} \sim L_{p}$.

 Apparently, this may be due to
the fact that QMFL with GUR (\ref{U2}) is unitary non-equivalent
to  QM with UR. Actually, in QM the left-hand side of (\ref{U6})
can be chosen arbitrary close to zero, whereas in QMFL this is
impossible. But if two theories are unitary equivalent, the form
of their traces should be retained. Besides, a more important
aspect is contributing to unitary non-equivalence of these two
theories: QMFL contains three fundamental constants (independent
parameters) $G$, $c$ and $\hbar$, whereas QM contains only one
$\hbar$. Within an inflation model \cite{r9}, QM is the limit of
QMFL (QMFL turns to QM) for the expansion of the Universe and low
energy limit. In this case the second term in the right-hand side
of (\ref{U2}) vanishes and GUR turn to UR. A natural way for
studying QMFL is to consider it as a deformation of QM, which
turns to the last one at the low energy limit (during the
Universe's expansion after the Big Bang). We will consider
precisely this option. However differing from authors of papers
\cite{r4},\cite{r5} and others we will deformed not commutators,
but density matrix, leaving at the same time the fundamental
measure quantum mechanical rule (\ref{U6}) without changes. Here
the following question may be formulated: how should be deformed
density matrix conserving Quantum Mechanics' measuring rules in
order to obtain self-consistent measuring procedure in QMFL? For
answering to the question we will use the wave packet formalism.
For starting let's consider a wave packet moving from Planck's
energy region to low energy one. Then the initial measurement is
of the order of Planck's scale $a \approx il_{min}$ or $a \sim
iL_{p}$. Further $a$ tends to infinity and we obtain for density
matrix $$Sp[\rho a^{2}]-Sp[\rho a]Sp[\rho a] \simeq
l^{2}_{min}\;\; or\;\; Sp[\rho]-Sp^{2}[\rho] \simeq
l^{2}_{min}/a^{2}.$$

 Therefore:

 \begin{enumerate}
 \item When $a < \infty$, $Sp[\rho] =
Sp[\rho(a)]$,
 $Sp[\rho]-Sp^{2}[\rho]>0$, and then \newline $Sp[\rho]<1$.
 This corresponds to the QMFL case.
\item When $a = \infty$, $Sp[\rho]$ does not depend on $a$,
$Sp[\rho]-Sp^{2}[\rho]\rightarrow 0$, and then $Sp[\rho]=1$. This
corresponds to the QM case.
\end{enumerate}

 Hence the  properties of density matrix in these two
theories have to be different. The only reasoning in this case may
be as follows: QMFL must differ from QM, but in such a way that in
the low energy limit a density matrix in QMFL be coincident with
the density matrix in QM. That is to say, QMFL is a deformation of
QM and the parameter of deformation depends on the measuring
scale. This means that in QMFL $\rho=\rho(x)$, where $x$ is the
scale, and for $x\rightarrow\infty$  $\rho(x) \rightarrow
\widehat{\rho}$, where $\widehat{\rho}$ is the density matrix in
QM.

Since on the Planck's scale $Sp[\rho]<1$, then for such scales
$\rho=\rho(x)$, where $x$ is the scale, is not a density matrix as
it is generally defined in QM. On Planck's scale we name $\rho(x)$
a "density pro-matrix". As follows from the above, the density
matrix $\widehat{\rho}$ appears in the limit
\begin{equation}\label{U12}
\lim\limits_{x\rightarrow\infty}\rho(x)\rightarrow\widehat{\rho},
\end{equation}
when GUR (\ref{U2}) turn to UR  and QMFL turns to QM.

Thus, on Planck's scale the density matrix is inadequate to obtain
all information about the mean values of operators. A "deformed"
density matrix (or pro-matrix) $\rho(x)$ with $Sp[\rho]<1$ has to
be introduced because a missing part of information $1-Sp[\rho]$
is encoded in the quantity $l^{2}_{min}/a^{2}$, whose specific
weight decreases as the scale $a$ expressed in  units of $l_{min}$
is going up.

\section{QMFL as a deformation of QM}
Here we describe QMFL as a deformation of QM using the
above-developed formalism of density pro-matrix. Within this
formalism, the density pro-matrix should be understood as a
deformed density matrix in QMFL. As a fundamental parameter of
deformation we use the quantity $\beta=l_{min}^{2 }/x^{2 }$, where
$x$ is the scale.

\noindent {\bf Definition 1.}

\noindent Any system in QMFL is described by a density pro-matrix
of the form $\rho(\beta)=\sum_{i}\omega_{i}(\beta)|i><i|$, where
\begin{enumerate}
\item $0<\beta\leq1/4$;
\item The vectors $|i>$ form a full orthonormal system;
\item $\omega_{i}(\beta)\geq 0$ and for all $i$  the
finite limit $\lim\limits_{\beta\rightarrow
0}\omega_{i}(\beta)=\omega_{i}$ exists;
\item
$Sp[\rho(\beta)]=\sum_{i}\omega_{i}(\beta)<1$,
$\sum_{i}\omega_{i}=1$;
\item For every operator $B$ and any $\beta$ there is a
mean operator $B$ depending on  $\beta$:\\
$$<B>_{\beta}=\sum_{i}\omega_{i}(\beta)<i|B|i>.$$
\end{enumerate}
Finally, in order that our definition 1  agree with the result of
section 2, the following condition must be fulfilled:
\begin{equation}\label{U13}
Sp[\rho(\beta)]-Sp^{2}[\rho(\beta)]\approx\beta.
\end{equation}
Hence we can find the value for
$Sp[\rho(\beta)]$ satisfying the condition of definition 1:
\begin{equation}\label{U14}
Sp[\rho(\beta)]\approx\frac{1}{2}+\sqrt{\frac{1}{4}-\beta}.
\end{equation}

According to point 5),  $<1>_{\beta}=Sp[\rho(\beta)]$. Therefore
for any scalar quantity $f$ we have $<f>_{\beta}=f
Sp[\rho(\beta)]$. In particular, the mean value
$<[x_{\mu},p_{\nu}]>_{\beta}$ is equal to
\begin{equation}\label{U15}
<[x_{\mu},p_{\nu}]>_{\beta}= i\hbar\delta_{\mu,\nu}
Sp[\rho(\beta)].
\end{equation}
We denote the limit $\lim\limits_{\beta\rightarrow
0}\rho(\beta)=\rho$ as the density matrix. Evidently, in the limit
$\beta\rightarrow 0$ we return to QM.

As follows from definition 1,
$<(j><j)>_{\beta}=\omega_{j}(\beta)$, from whence the completeness
condition by $\beta$ is
\\$<(\sum_{i}|i><i|)>_{\beta}=<1>_{\beta}=Sp[\rho(\beta)]$. The
norm of any vector $|\psi>$ assigned to  $\beta$ can be defined as
\\$<\psi|\psi>_{\beta}=<\psi|(\sum_{i}|i><i|)_{\beta}|\psi>
=<\psi|(1)_{\beta}|\psi>=<\psi|\psi> Sp[\rho(\beta)]$, where
$<\psi|\psi>$ is the norm in QM, i.e. for $\beta\rightarrow 0$.
Similarly, the described theory may be interpreted using a
probabilistic approach, however requiring  replacement of $\rho$
by $\rho(\beta)$ in all formulae.

\renewcommand{\theenumi}{\Roman{enumi}}
\renewcommand{\labelenumi}{\theenumi.}
\renewcommand{\labelenumii}{\theenumii.}

It should be noted:

\begin{enumerate}
\item The above limit covers both Quantum
and Classical Mechanics. Indeed, since $\beta\sim L_{p}^{2 }/x^{2
}=G \hbar/c^3 x^{2 }$, we obtain:
\begin{enumerate}
\item $(\hbar \neq 0,x\rightarrow
\infty)\Rightarrow(\beta\rightarrow
0)$ for QM;
\item $(\hbar\rightarrow 0,x\rightarrow
\infty)\Rightarrow(\beta\rightarrow
0)$ for Classical Mechanics;
\end{enumerate}
\item As a matter of fact, the deformation parameter $\beta$
should assume the value $0<\beta\leq1$.  However, as seen from
(\ref{U14}), $Sp[\rho(\beta)]$ is well defined only for
$0<\beta\leq1/4$. Some problems can be associated with the point,
where $\beta=1/4$. If $x=il_{min}$ and $i\geq 2$, this problem
vanishes. At the point where $x=l_{min}$ there is a singularity
related to complex values assumed by $Sp[\rho(\beta)]$ , i.e. to
the impossibility of obtaining a diagonalized density pro-matrix
at this point over the field of real numbers. For this reason
definition 1 has no sense at the point $x=l_{min}$. We will come
back to the question appearing in this section when we will
discuss singularity and hypothesis of cosmic censorship in section
5.

\item We consider possible solutions for (\ref{U13}).
For instance, one of the solutions of (\ref{U13}), at least to the
first order in $\beta$, is $$\rho^{*}(\beta)=\sum_{i}\alpha_{i}
exp(-\beta)|i><i|,$$ where all $\alpha_{i}>0$ are independent of
$\beta$  and their sum is equal to 1. In this way
$Sp[\rho^{*}(\beta)]=exp(-\beta)$. Indeed, we can easily verify
that \begin{equation}\label{U15}
Sp[\rho^{*}(\beta)]-Sp^{2}[\rho^{*}(\beta)]=\beta+O(\beta^{2}).
\end{equation}
Note that in the momentum representation
$\beta=p^{2}/p^{2}_{max}$, where $p_{max}\sim p_{pl}$ and $p_{pl}$
is the Planck momentum. When present in matrix elements,
$exp(-\beta)$ can damp the contribution of great momenta in a
perturbation theory.
\item It is clear that within the proposed description the
states with a unit probability, i.e. pure states,
can appear only in the limit $\beta\rightarrow 0$,
when all $\omega_{i}(\beta)$ except for one are equal
to zero or when they tend to zero at this limit.
In our treatment pure state are states, which can be represented
in the form $|\psi><\psi|$, where $<\psi|\psi>=1$.

\item We suppose that all the definitions concerning a
density matrix can be transferred to the above-mentioned
deformation of Quantum Mechanics (QMFL) through changing the
density matrix $\rho$ by the density pro-matrix $\rho(\beta)$ and
subsequent passage to the low energy limit $\beta\rightarrow 0$.
Specifically, for statistical entropy we have
\begin{equation}\label{U16}
S_{\beta}=-Sp[\rho(\beta)\ln(\rho(\beta))].
\end{equation}
The quantity of $S_{\beta}$ seems never to be equal to zero as
$\ln(\rho(\beta))\neq 0$ and hence $S_{\beta}$ may be equal
to zero at the limit $\beta\rightarrow 0$ only.
\end{enumerate}
\newpage
Some Implications:
\begin{enumerate}
\item If we carry out measurement on the pre-determined scale, it is
impossible to regard the density pro-matrix as a density matrix with an
accuracy better than particular limit $\sim10^{-66+2n}$, where
$10^{-n}$ is the measuring scale. In the majority of known cases
this is sufficient to consider the density pro-matrix as a density
matrix. But on Planck's scale, where the quantum gravitational
effects and Plank energy levels cannot be neglected, the difference
between $\rho(\beta)$ and  $\rho$ should be taken into consideration.

\item Proceeding from the above, on Planck's scale the
notion of Wave Function of the Universe (as introduced in
\cite{r10}) has no sense, and quantum gravitation effects in this
case should be described with the help of density pro-matrix
$\rho(\beta)$ only.
\item Since density pro-matrix $\rho(\beta)$ depends on the measuring
scale, evolution of the Universe within the inflation model
paradigm \cite{r9} is not a unitary process, or otherwise the
probabilities $p_{i}=\omega_{i}(\beta)$  would be preserved.
\end{enumerate}

\section{Dynamical aspects of QMFL}
Let's suppose that in QMFL density pro-matrix has the form
$\rho[\beta(t),t]$, or in other words it depends on two
parameters: time $t$ and parameter of deformation $\beta$, which
also depends on time $\beta=\beta(t)$. Then we have
\begin{equation}\label{U17}
\rho[\beta(t),t]=\sum\omega_{i}[\beta(t)]|i(t)><i(t)|.
\end{equation}
Differentiating the last expression on time we obtain the equation
\begin{equation}\label{U18}
\frac{d\rho}{dt}=\sum_{i}
\frac{d\omega_{i}[\beta(t)]}{dt}|i(t)>-i[H,\rho(\beta)]=d[ln\omega(\beta)]\rho
(\beta)-i[H,\rho(\beta)].
\end{equation}
Where $ln[\omega(\beta)]$ is a row-matrix and $\rho(\beta)$ is a
column-matrix. Thus we obtain a prototype of the
 Liouville's equation.

Let's consider some particular cases.
\begin{enumerate}
\item First we consider the process of time
evolution at low energies, or in other words, when $\beta(0)
\approx 0$, $\beta(t)\approx 0$ and $t \to \infty$. Then it is
clear that $\omega_{i}(\beta)\approx \omega_{i} \approx constant$.
The first term in (\ref{U18}) vanishes and we obtain the
Liouville's equation.
\item We obtain also the Liouville's equation when we turn from
inflation to big scale. Within the inflation approach the scale $a
\approx e^{Ht}$, where $H$ is the Hubble's constant and $t$ is
time. Therefore $\beta \sim e^{-2Ht}$ and when $t$ is quite big
$\beta \to 0$. In other words $\omega_{i}[\beta] \to \omega_{i}$,
the first term in (\ref{U18}) vanishes and we again obtain the
Liouville's equation.
\item At very early stage of inflation process or even before it
takes place $\omega_{i}[\beta]$ was not a constant and therefore
the first term in (\ref{U18}) should be taking into account. This
way we obtain a deviation from the Liouville's equation.
\item At last let's consider the case when $\beta(0) \approx 0$,
$\beta(t)>0$ when $t \to \infty$. In this case we are going from
low energy scale to high one and $\beta(t)$ grows when $t \to
\infty$. In this case the first term in (\ref{U18}) is significant
and we obtain an addition to the Liouville's equation in the form
$$d[ln\omega(\beta)]\rho(\beta).$$ This case could take place when
matter go into a black hole and is moving in direction of the
singularity (to the Planck's scale).
\end{enumerate}

 \section{Singularity, Entropy and Information Loss in Black
Holes} It follows to note that remark II in section 3 about
complex meaning assumed by density pro-matrix at the point with
fundamental length has straightforward relation with the
singularity problem and cosmic censorship in General Theory of
Relativity \cite{r11}. Indeed, considering, for instance, a
Schwarzchild's black hole (\cite{r12}, p.9) with metrics:
\begin{equation}\label{U19}
 ds^2 = - (1 - \frac{2M}{r}) dt^2 +
\frac{dr^2}{(1 - \frac{2M}{r})} + r^2 d \Omega_{II}^2,
\end{equation}
we obtain, as it is well-known a singularity at the point $r=0$.
In our approach this corresponds to the point with fundamental
length ($r=l_{min}$). At this point we are not able to measure
anything, since at this point $\beta=1$ and $Sp[\rho (\beta)]$
becomes complex. Thus, we carry out a measurement, starting from
the point $r=2l_{min}$ corresponding to $\beta=1/4$. Consequently,
the initial singularity $r=l_{min}$, which cannot be measured, is
hidden of observation. This confirms the hypothesis of cosmic
censorship in this concrete case. This hypothesis claims that "a
naked singularity cannot be observed". Thus QMFL in our approach
feels the singularity. (In comparison with QM, which does not feel
it).

Statistical entropy, connected with density pro-matrix and
introduced in the remark V section 3
$$S_{\beta}=-Sp[\rho(\beta)\ln(\rho(\beta))]$$  may be interpreted
as density entropy on unity of minimal square $l^{2}_{min}$
depending on the scale $x$. It could be quite big nearby the
singularity. In other words, when $\beta\rightarrow 1/4$. This
does not contradict the second law of Thermodynamics since the
maximal entropy of a determined object in the Universe is
proportional to the square of their surface $A$, measured in units
of minimal square $l^{2}_{min}$ or Planck's square $L_{p}^2$, as
it was shown in some papers (see for instance \cite{r13}).
Therefore, in the expanded Universe since surface $A$ grows, then
entropy does not decrease.

The obtained results enable one to consider anew the information
loss problem associated with black holes \cite{r14,r15}, at least
for the case of primordial ones. Indeed, when we consider the
black holes, Planck's scale is a factor. And it was shown that the
entropy of matter absorbed by a black hole on this scale is not
equal to zero, supporting the data of R.Myers \cite{r16}.
According to his results, a pure state cannot form a black hole.
Then  it is necessary to reformulate the problem per se, since so
far in all the publications on information paradox zero entropy of
the initial state has been compared to  nonzero entropy of the
final state. Besides, it should be noted that in some recent
papers for all types of black holes QM with GUR is considered at
the very beginning \cite{r17}. As a consequence of this approach,
stable remnants with masses of the order of Planck's ones $M_{pl}$
emerge during the process of black hole evaporation. From here it
follows that black holes should not evaporate fully. We arrive to
the conclusion that results given in \cite{r12, r18} are correct
only in the semi-classical approximation and they should not be
applicable to the quantum back hole analysis.

 Based on our results, we
can elucidate (at least qualitatively) the problem associated with
information loss on black holes formed when a star collapses.
Actually, near the horizon of events the entropy of an
approximately pure state is practically equal to zero:
$S^{in}=-Sp[\rho \ln(\rho)]$ that is associated with the value
$\beta \mapsto 0$. When approaching the singularity $\beta>0$
(i.e. on Plank's scale), its entropy is nonzero for
$S_{\beta}=-Sp[\rho (\beta)\ln(\rho(\beta))]$. Therefore, in such
a black hole the entropy increases, whereas information is lost.

On the other hand, from the results obtained above, at least at
the qualitative level, it can be clear up the answer to the
question how may be information lost in big black holes, which are
formed as result of star collapse. Our point of view is closed to
the R. Penrose's one \cite{r19}. He considers (in opposition to S.
Hawking) that information in black holes is lost when matter meets
a singularity. In our approach information loss takes place in the
same form. Indeed, near to the horizon of events an approximately
pure state with practically equal to zero initial entropy
$S^{in}=-Sp[\rho\ln(\rho)]$, which corresponds to $\beta \to 0$,
when approaching a singularity (in other words is reaching the
Planck's scale) gives yet non zero entropy
$S_{\beta}=-Sp[\rho(\beta)\ln(\rho(\beta))]>0$ when $\beta >0$.
Therefore, entropy increases and information is lost in this black
hole. We can (at the moment, also at the qualitative level)
evaluate entropy of black holes. Indeed, starting from density
matrix for a pure state at the "entry" of a black hole
$\rho_{in}=\rho_{pure}$ with zero entropy $S^{in}=0$, we obtain,
doing a straightforward calculation
 at
the singularity in the black hole density pro-matrix
$\rho_{out}=\rho(\beta)\approx \rho(1/4)$ with entropy
$$S^{out}=S_{1/4}=-Sp[\rho(1/4)\ln(\rho(1/4)]= -1/2 \ln1/2 \approx
0.34657.$$
 Taking into account that total entropy of a
black hole is proportional to quantum area of surface A, measured
in Planck's units of area $L_{p}^2$ \cite{r20}, we obtain the
following value for quantum entropy of a black hole:
\begin{equation}\label{U20}
S_{black  hole}\sim 0.34657 A.
\end{equation}

This formula differs from the well-known one given by
Bekenstein-Hawking for black hole entropy $S_{black
hole}=\frac{1}{4} A$ \cite{r21}. This result was obtained in the
semi-classical approximation. At the present moment quantum
corrections to this formula are object of investigation
\cite{r22}. As it was yet above-mentioned we carry out a
straightforward calculation. Namely, using the anzats of the
remark III in section 3 and assuming the density pro-matrix equal
to $Sp[\rho^{*}(\beta)]=exp(-\beta)$, we obtain for $\beta=1/4$
that entropy is equal to $$S^{* out}=S^{*}_{1/4}=-Sp[exp(-1/4)\ln
exp(-1/4)]\approx 0.1947,$$ and consequently we arrive to the
value for entropy
\begin{equation}\label{U21}
S_{black  hole} \sim 0.1947 A
\end{equation}
which is in good agreement with the result given in  \cite{r22}.
Our approach, leading to formula (\ref{U21}) is at the very
beginning,
 based on the quantum nature of
black holes. Let us to note here, that in the approaches, used up
to now to modify Liouville's equation, due to information paradox
\cite{r23}, the added member appearing in the right side of
(\ref{U18}) has the form $$-\frac{1}{2}\sum_{\alpha \beta \neq 0}
(Q^{\beta}Q^{\alpha}\rho+\rho Q^{\beta}Q^{\alpha}-2 Q^{\alpha}\rho
Q^{\beta}),$$ where $Q^{\alpha}$
  is a full orthogonal set
of Hermitian matrices with $Q^{0} =1$. In this case either
locality or conservation of energy-impulse tensor is broken down.
In the offered in this paper approach, the added member in the
deformed Liouville's equation  has a more natural and beautiful
form in our opinion: $$d[ln\omega(\beta)]\rho (\beta).$$ In the
limit $\beta\to 0$ all properties of QM are conserved, the added
member vanishes and we obtain Liouville's equation.

\section{Some comments on Shr{\"o}dinger's picture}
A procedure allowing to obtain a theory from the transformation of
the precedent one is named "deformation". This is doing, using one
or a few parameters of deformation in such a way, that the
original theory must appear in the limit, when all parameters tend
to some fixed values. The most clear example is QM being a
deformation of Classical Mechanics. The parameter of deformation
in this case is the Planck's constant $\hbar$. When
$\hbar\rightarrow 0$ QM passages to Classical Mechanics. As it was
indicated above in the remark 1 section 3, we are able to obtain
from QMFL two limits: Quantum and Classical Mechanics. The
described here deformation should be understood as "minimal" in
the sense that we have deformed only the probability
$\omega_{i}\rightarrow \omega_{i}(\beta)$, whereas state vectors
have been not deformed. In a most complete consideration we will
be obligated to consider instead $|i><i|$, vectors
$|i(\beta)><i(\beta)|$ and in this case the full picture will be
very complicated. It is easy to understand how Shrodinger's
picture is transformed in QMFL. The prototype of Quantum
Mechanical normed wave function $\psi(q)$ with
$\int|\psi(q)|^{2}dq=1$ in QMFL is $\theta(\beta)\psi(q)$. The
parameter of deformation $\beta$ assumes the value
$0<\beta\leq1/4$. Its properties are
$|\theta(\beta)|^{2}<1$,$\lim\limits_{\beta\rightarrow
0}|\theta(\beta)|^{2}=1$ and the relation
$|\theta(\beta)|^{2}-|\theta(\beta)|^{4}\approx \beta$ takes
place. In such a way the full probability always is less than 1:
$p(\beta)=|\theta(\beta)|^{2}=\int|\theta(\beta)|^{2}|\psi(q)|^{2}dq<1$
and it tends to 1 when  $\beta\rightarrow 0$. In the most general
case of arbitrarily normed state in QMFL
$\psi=\psi(\beta,q)=\sum_{n}a_{n}\theta_{n}(\beta)\psi_{n}(q)$ c
$\sum_{n}|a_{n}|^{2}=1$ the full probability is
$p(\beta)=\sum_{n}|a_{n}|^{2}|\theta_{n}(\beta)|^{2}<1$ and
 $\lim\limits_{\beta\rightarrow 0}p(\beta)=1$.

 It is natural that in QMFL Shrodinger's equation is also
deformed. It is replaced by the  equation
\begin{equation}\label{U22}
\frac{\partial\psi(\beta,q)}{\partial t}
=\frac{\partial[\theta(\beta)\psi(q)]}{\partial
t}=\frac{\partial\theta(\beta)}{\partial
t}\psi(q)+\theta(\beta)\frac{\partial\psi(q)}{\partial t},
\end{equation}
where the second term in the right side generates the Shrodinger's
equation since
\begin{equation}\label{U23}
\theta(\beta)\frac{\partial\psi(q)}{\partial
t}=\frac{-i\theta(\beta)}{\hbar}H\psi(q).
\end{equation}

Here $H$ is the Hamiltonian and the first member is added,
similarly to the member appearing in the deformed Loiuville's
equation and  vanishing when $\theta[\beta(t)]\approx const$. In
particular, this takes place in the low energy limit in QM, when
$\beta\rightarrow 0$.  It follows to note that the above-described
theory  is not time-reversal as QM, since the combination
$\theta(\beta)\psi(q)$ breaks down this property in the deformed
Shrodinger's equation. Time-reversal is conserved only in the low
energy limit, when quantum mechanical Shrodinger's equation is
valid.

\section{Conclusion}
It follows to note, that in some well-known papers on GUR and
Quantum Gravity (see for instance \cite{r1,r2,r3,r4,r24}) there is
not any mention about the measuring procedure. However, it is
clear that this question is crucial and it cannot be ignored or
passed over in silence. Taking into account this state of affairs
we propose in this paper a detailed tratment of this problem. In
the present paper the measuring rule (\ref{U6}) is proposed as a
good initial approximation to the exact measuring procedure in
QMFL. Corrections to this procedure could be defined by an
adequate and fully established description of the space-time foam
(see \cite{r25}) on Planck's scale. On the other hand, as it was
noted in  (see \cite{r26}) all known approaches
 dealing with Quantum Gravity
one way or another lead to the notion of fundamental length .
Involving that notion too, GUR (\ref{U2}) are well described by
the inflation model \cite{r27}. Therefore, it seems impossible to
understand physics on Planck's scale disregarding the notion of
fundamental length. One more aspect of this problem should be
considered. As it was noted in \cite{r28}, advancement of a new
physical theory implies the introduction of a new parameter and
deformation of the precedent theory by this parameter. In essence,
all these deformation parameters are fundamental constants: $G$,
$c$ and $\hbar$ (more exactly in \cite{r28} $1/c$ is used instead
of $c$). As follows from the above results, in the problem from
\cite{r28} one may redetermine, whether a theory we are seeking is
the theory with fundamental length involving these three
parameters by definition: $L_{p}=\sqrt\frac{G\hbar}{c^3}$. Notice
also that the deformation introduced in this paper is stable in
the sense indicated in \cite{r28}.

\noindent In the present paper the approach first developed in
\cite{r29} is improved.


\end{document}